\documentclass[12pt]{article}
\usepackage{graphicx}
\usepackage{amsmath}

\title{ Reconstruction of $\tilde{\tau}_{1}$ mass at the LHC}

\author{\textbf{Rashid M. Djilkibaev$^{1}$ \thanks{Permanent address:
     Institute for Nuclear Research, 60-th Oct. pr. 7a,
     Moscow 117312, Russia}\ ,
     Rostislav V. Konoplich$^{1,2}$}\\
   \normalsize$^{1}$Department of Physics, New York University,
   New York, NY 10003\\
   \normalsize$^{2}$Manhattan College, Riverdale, New York, NY, 10471}

\begin{document}

\maketitle

\begin{abstract}
The cascade mass reconstruction approach  was used for 
mass reconstruction of the lightest stau produced at the LHC 
in the cascade decay $\tilde{g} \to \tilde{b} b \to \tilde{\chi}_{2}^{0} 
b b \to \tilde{\tau}_{1} \tau b b \to \tilde{\chi}_{1}^{0} \tau \tau b b $. 
The stau mass was reconstructed assuming that masses of 
gluino, bottom squark and two lightest neutralinos were
reconstructed in advance.

SUSY data sample sets for the 
SU3 model point containing 160k events each were generated
which corresponded to an integrated luminosity of about 
$8 \rm fb^{-1}$ at 14 TeV. 
These  events were passed through the AcerDET detector
simulator, which parametrized the response of 
a generic LHC detector.
The mass of the $\tilde{\tau}_{1}$ was reconstructed with 
a precision of about $20\%$ on average.

\end{abstract}

\newpage

\section*{I. Introduction} 

 If supersymmetry exists at an energy scale of ~1 TeV, 
SUSY particles such as gluinos and squarks should be abundantly 
produced at the LHC. Assuming R-parity conservation,
these particles cascade down to the lightest supersymmetric
particles (LSPs). 
A detailed discussion of possible SUSY effects at the LHC is given in \cite{atlas}. 

   In this paper we consider a mass reconstruction of the lightest stau ($\tilde{\tau}_{1}$)
in the cascade decay

\begin{equation}
\tilde{g} \to \tilde{b} b \to \tilde{\chi}_{2}^{0} b b \to \tilde{\tau}_{1} \tau b b \to \tilde{\chi}_{1}^{0} \tau \tau b b 
\label{chain} 
\end{equation}

\begin{figure}[h]
  \centering
  \includegraphics[width=0.7\textwidth]{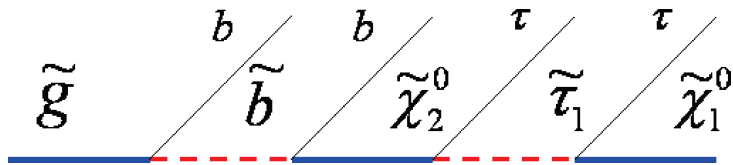} 
  \caption[Short caption.]{A gluino cascade decay chain.}
\label{fig:chain}
\end{figure}

The gluino decay chain (\ref{chain}) is shown in Fig.(\ref{fig:chain})

The study of sleptons and squarks of third generation is of a special interest.
Their masses can be very different than that of sparticles of the first and second 
generation, because of the effects of large Yukawa and soft couplings
in the renormalization group equations. Furthermore they can show large mixing in
pairs $(\tilde{t}_{L}, \tilde{t}_{R}), (\tilde{b}_{L}, \tilde{b}_{R})$
and $(\tilde{\tau}_{L}, \tilde{\tau}_{R})$. 
The properties of $\tilde{\tau}_{1}$ are also
important for determination of the dark matter relic density.

The reconstruction of a SUSY event is complicated because of 
escaping LSPs and many complex and competing decay modes.
At present there are two different approaches to SUSY mass reconstruction.
The ``endpoint method", which has been widely studied 
\cite{baer} - \cite{lester} for the LHC at high integrated
luminosity of about $100-300 ~\rm fb^{-1}$, looks 
for kinematic endpoints of invariant mass distributions.
The second method of SUSY particle mass reconstruction  
is the ``mass relation approach" \cite{tovey} - \cite{nojiri}, 
based on the mass relation 
equation which relates SUSY particle masses and 
measured momenta of detected particles. It was shown in \cite{dk}
that the "mass relation approach" can be used for luminosities
as low as a few $fb^{-1}$.  

In this work the cascade mass reconstruction approach \cite{dk}, 
that we developed earlier and applied to a different SUSY
cascade decay, is used for 
mass reconstruction of the lightest stau at the LHC energy of 14 TeV 
with an integrated luminosity of about $8 ~\rm fb^{-1}$.
The stau mass is reconstructed assuming that the masses of 
the gluino, bottom squark and two lightest neutralinos were
reconstructed in advance.
At high integrated luminosity the stau mass could be extracted 
directly from the endpoint of $\tau \tau$ invariant mass distribution.
However at integrated luminosities below 10 $\rm fb^{-1}$ 
the problem becomes quite challenging because
there is no a sharp edge in the $\tau \tau$ invariant mass distribution
because of escaping neutrinos, a reconstructed stau mass is very
sensitive to uncertainties in input parameters (neutralino masses)
and there is a high level of SUSY background.

The cascade mass reconstruction approach \cite{dk} applied for 
stau mass reconstruction in this work is based on a consecutive use of the endpoint method,
an event filter and a combinatorial mass reconstruction method. 
The endpoint method is used to get a rough estimate of the
stau mass and the corresponding errors.
This first estimate of mass is used by
an event filter for each event in 
the data sample to reduce background for the following 
mass reconstruction step.
Finally, the stau mass is reconstructed by a 
maximization of a combined likelihood function,  
which depends on all five sparticle masses
(gluino, bottom squark, stau and two lightest neutralinos), and is
constructed for each possible
combination of five events in the data sample. 

\section*{II. Simulation }

We choose for this study the SU3 model point. 
This point has a significant production cross section for the 
chain (\ref{chain}); gluinos and squarks should be produced
abundantly at the LHC.
The bulk point SU3 
is the official benchmark point of the ATLAS collaboration 
and it is in agreement with
the recent precision WMAP data ~\cite{wmap}. 
This model point is described by the
set of mSUGRA parameters given in Table (\ref{tab:paramth}).

\begin{table} [h]
\begin{center}
 \begin{tabular}{|c|c|c|c|c|c|}
   \hline
   Point & $m_{0}$ & $m_{1/2}$ & $A_{0}$ & $tan\beta$ & $\mu$ \\
   \hline\hline
   SU3 & 100 GeV & 300 GeV & -300 GeV & 6 &  $>$ 0 \\
   \hline
 \end{tabular}
\caption{mSUGRA parameters for the SU3 point.}\label{tab:paramth}
\end{center}
\end{table} 

Assumed theoretical masses of SUSY particles in the cascade (\ref{chain}), the total branching ratio  
and a cross section
generated by ISAJET 7.74 \cite{isajet} are given in Table (\ref{tab:massth}).

\begin{table} [h]
\begin{center}
 \begin{tabular}{|c|c|c|c|c|c|c|c|c|}
   \hline
   Point & $m_{\tilde{g}}$ & $m_{\tilde{b_{1}}}$ & $m_{\tilde{b}_{2}}$ & $m_{\tilde{\chi}_{2}^{0}}$ & 
   $m_{\tilde{tau}_{1}}$ & $m_{\tilde{\chi}_{1}^{0}}$ & BR& $\sigma [\rm pb]$ \\
   \hline\hline
   SU3 & 720.16 & 605.93 & 642.00 & 223.27 & 151.46 & 118.83 & 2.71$\%$ &19\\
   \hline
 \end{tabular}
\caption{Assumed  theoretical masses of sparticles, branching ratio BR and production cross section $\sigma $ 
at the SU3 point. 
Masses are given in GeV.}\label{tab:massth}
\end{center}
\end{table}  

Branching ratios for the gluino decay chain (\ref{chain}) at the SU3 point are

\begin{equation}
\nonumber
\tilde{g} \stackrel{16.6 \%}{\longrightarrow} \tilde{b}_{1} \stackrel{24.1 \%}{\longrightarrow} \tilde{\chi}_{2}^{0} 
\stackrel{48.7 \%}{\longrightarrow} \tilde{\tau}_{1} \stackrel{100 \%}{\longrightarrow} \tilde{\chi}_{1}^{0} ~~\Rightarrow ~~1.96 \% 
\end{equation}

\begin{equation}
\nonumber
\tilde{g} \stackrel{9.2 \%}{\longrightarrow} \tilde{b}_{2} \stackrel{16.6 \%}{\longrightarrow} \tilde{\chi}_{2}^{0} 
\stackrel{48.7 \%}{\longrightarrow}  \tilde{\tau}_{1} \stackrel{100 \%}{\longrightarrow} \tilde{\chi}_{1}^{0} ~~\Rightarrow ~~0.75 \% 
\end{equation}

\noindent

Monte Carlo simulations of SUSY production at model points were performed by the HERWIG 6.510
event generator  ~\cite{herwig}. The produced events were passed through the AcerDET 
detector simulation ~\cite{atlfast}, which parametrized the response of a 
detector (LHC detector descriptions can be found in \cite{det_atlas}, \cite{det_cms}). 
The efficiency for jet reconstruction and labeling was $80\%$.
An additional factor of $50\%$ took into account tau efficiency, which
typically corresponded to a rejection factor of about 100 for QCD jets background. 
Samples of 160k  SUSY events were used. 
This approximately  corresponds  to $8 ~\rm fb^{-1}$ of integrated luminosity
for the SUSY SU3 point production cross section of 19 pb at 14 TeV.
Five different sets of 160k  SUSY events 
were considered to demonstrate the stability and precision of the mass reconstruction 
approach.

As input parameters for the stau mass reconstruction procedure, 
hypothetical masses of the gluino, bottom squark and two lightest neutralinos
are used. These input masses are assumed to have been 
reconstructed by the cascade mass reconstruction approach 
for SUSY data sets corresponding to $4 ~\rm fb^{-1}$ \cite{dk} and are presented 
in Table \ref{tab:reco1}.

\begin{table} [htb!]
\begin{center}
 \begin{tabular}{|c|c|c|c|c|}
   \hline
   Set &$m_{\tilde{g}}$ & $m_{\tilde{b}}$ & $m_{\tilde{\chi}_{2}^{0}}$ & $m_{\tilde{\chi}_{1}^{0}}$\\
   \hline
   1&701$\pm$57 &600$\pm$57 &208$\pm$21 &98$\pm$22 \\
   2&712$\pm$55 &608$\pm$53 &254$\pm$21 &143$\pm$20 \\
   3&664$\pm$78 &564$\pm$80 &219$\pm$24 &109$\pm$23 \\
   4&767$\pm$62 &649$\pm$65 &258$\pm$35 &148$\pm$34 \\
   5&655$\pm$45 &545$\pm$47 &208$\pm$21 &96$\pm$20 \\
   \hline
 \end{tabular}
\caption{
Reconstructed SUSY particle masses and reconstruction errors for five data sample sets.
}
\label{tab:reco1}
\end{center}
\end{table} 

In order to isolate the chain (\ref{chain}) the following cuts were applied:

$\bullet$ two jets tagged as $\tau$ with opposite charge
satisfying transverse momentum cuts $p_{T} > 30~GeV$ and 
$p_{T} >  25~GeV$

$\bullet$ two b-tagged jets; 

$\bullet$ at least four jets, satisfying 
$p_{T1} > 100~GeV$, $p_{T2} > 50~GeV$, $p_{T3} > 50~GeV$, $p_{T4} > 50~GeV$;  

$\bullet$ $M_{eff} > 500~GeV$ and $E_{T}^{miss} > 0.2M_{eff}$, where
$E_{T}^{miss}$ is the event's missing transverse energy and $M_{eff}$ is the scalar
sum of the missing transverse energy and the transverse momenta of the four
hardest jets; 

$\bullet$ invariant mass of $\tau$-tagged jets satisfying $10~GeV < M_{\tau \tau} < 300~GeV$.

Note that at the first stage of the reconstruction procedure, the stau mass is  
estimated by considering the chain
$\tilde{q_L} \to \tilde{\chi}_{2}^{0} q \to \tilde{\tau} \tau_{2} q \to \tilde{\chi}_{1}^{0} \tau \tau q$
with the same cuts as mentioned above except
requiring b-tagged jets.  

It was shown in \cite{gj_note} that the Standard Model processes are suppressed 
significantly by the above requirements on $\tau$-tagged jets. The Standard Model 
dominant backgrounds 
surviving the hard cuts are Z + jets and  $t\bar{t}$ production, where
both W's decay leptonically into a $bbll$ state. This background
was estimated to be 1/10 of the SUSY backgrounds \cite{gj_note}. The $t\bar{t}$ background
can be reduced  
to about $2\%$ by applying the event filter procedure of the cascade mass
reconstruction approach \cite{dk}.
Thus, the remaining Standard Model contribution is negligible in comparison 
with that of SUSY backgrounds, and therefore Standard Model background
is not included in the following analysis, the SUSY
background is about three times the signal rate.

Table (\ref{tab:sigbcgsu3}) 
shows the number of signal events and SUSY
background events for the SU3 model point after cuts were applied to 
the five sets of 160k SUSY events. 
The dominant SUSY background consists of $\tau^+$ and $\tau^-$
produced in two different decay chains.
The classification of events as signal and SUSY background is based on 
simulated truth information.
The SUSY background to the process (\ref{chain})
is significant.
It follows from Table (\ref{tab:sigbcgsu3}) that for the SU3 point the number of SUSY 
background events is  a factor 3
greater than the number of signal events.

\begin{table} [htb!]
\begin{center}
 \begin{tabular}{|c|c|c|c|c|}
   \hline
   Set & Total & Signal & SUSY Backg. & Ratio  \\
   \hline\hline
   1 & 199 & 47 & 152 & 4.2\\
   2 & 212 & 56 & 156 & 3.8\\
   3 & 216 & 53 & 163 & 4.1\\
   4 & 216 & 48 & 168 & 4.5\\
   5 & 229 & 65 & 164 & 3.5\\
   \hline
   1-5 & 1072 & 269 & 803 & 4.0 \\
   \hline
 \end{tabular}
\caption{The number of signal and SUSY background events after cuts applied to
160k SUSY events. Ratio = (Signal+Background)/Signal.
Row 1-5 shows the number of events for five sets combined.}\label{tab:sigbcgsu3}
\end{center}
\end{table}

\section*{III. Preliminary estimate of stau mass by the endpoint method with low statistics}

The endpoint method has been widely used to determine masses of 
SUSY particles ~\cite{endpoints}, in particular it has been applied to the 
decay chain \cite{gj_note}, \cite{ibe} - \cite{peter}

\begin{equation}
\tilde{q_L} \to \tilde{\chi}_{2}^{0} q \to \tilde{\tau}_{1} \tau q \to \tilde{\chi}_{1}^{0} \tau \tau q
\label{chainqL} 
\end{equation}

\noindent
which is a subprocess of the cascade (\ref{chain}) if one considers
$\tilde{q_L}$ instead of $\tilde{b}$.

\begin{figure}[htb!]
  \centering
  \includegraphics[width=0.6\textwidth]{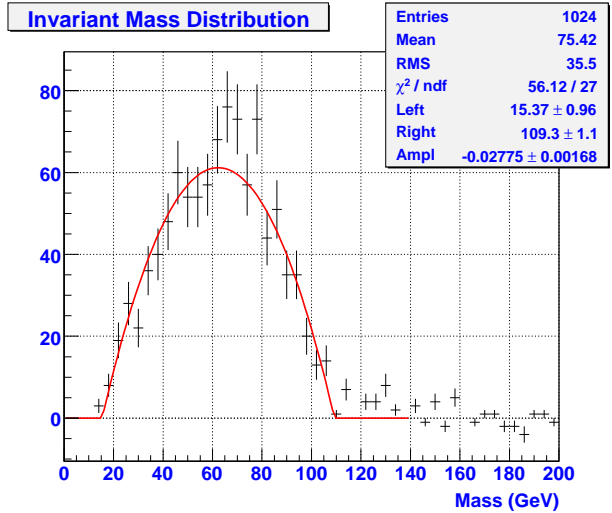} 
  \caption[Short caption.]{Invariant mass distribution of $\tau$-tagged jets for Set 1.} 
  \label{fig_edge1}
\end{figure}

Fig. (\ref{fig_edge1})  shows the $\tau \tau$ invariant mass 
distribution for Set 1 of 160k  events at the SU3 point
after application of the kinematic cuts and after 
subtraction of same sign $\tau$-tagged (background) jets was performed. This allowed
to reduce background because in background processes
positive and negative jets are mostly uncorrelated.

The distribution in Fig. (\ref{fig_edge1}) does not have a
sharp edge because of undetected neutrino energy in tau-decays
but the end-point is clearly seen at about 109 GeV and can be reconstructed.  

To extract endpoints from  the $\tau \tau$ invariant mass distribution 
a fit to this distribution was performed using a parabola plus a straight line
that matched the parabola. 

The results of the fit for the five data sets
are shown in Table (\ref{tab:endp}).

\begin{table} [htb!]
\begin{center}
 \begin{tabular}{|c|c|c|c|c|c|}
   \hline
    & Set 1 & Set 2 & Set 3 & Set 4 & Set 5  \\
   \hline\hline
   $endpoint$ & 109.3$\pm$1.1 & 105.0$\pm$1.0 & 104.0$\pm$0.7 & 108.5$\pm$1.0 & 106.8$\pm$0.6\\
   \hline
 \end{tabular}
\caption{Endpoints determined from fitting of $\tau \tau$-edges.}
\label{tab:endp}
\end{center}
\end{table} 

These results have to be compared with the theoretical endpoint 
of 101.7 GeV for the SU3 point. The fit was performed by MINUIT  \cite{minuit}
which gave errors presented in  Table (\ref{tab:endp}). These 
uncertainties are small because of a specific form of the fit function.
More conservative estimates of uncertainties \cite{gj_note}, \cite{atlnote}
are about 4 GeV but 
in our case this contribution is not important because the most
significant uncertainties come from $20\%$ uncertainties in sparticle masses.  

Once endpoints are found the stau mass
can be preliminary estimated from the analytical
expression for the $\tau \tau$ endpoint

\begin{equation} 
   m_{\tau \tau}^2 = (\tilde \xi - \tilde \tau)(\tilde \tau - \tilde \chi)/\tilde \tau
\label{endtau}
\end{equation}

 The following notations are used
for masses
$\tilde \chi = m_{\tilde \chi_{1}^{0}}^2$~, $\tilde \tau = m_{\tilde \tau_{1}}^2$~, 
$\tilde \xi = m_{\tilde \chi_{2}^{0}}^2$.

The results of the fit 
for preliminary stau mass estimates and their errors 
based on found 
endpoints, edge reconstruction errors, sparticle masses
and their errors 
are summarized in Table (\ref{tab:massfitsu3}).

\begin{table} [htb!]
\begin{center}
 \begin{tabular}{|c|c|c|c|c|c|}
   \hline
   $\tilde \tau_{1}$ & Set 1 & Set 2 & Set 3 & Set 4 & Set 5  \\
   \hline\hline
   solution 1 & 149$\pm$135 & 209$\pm$48 & 174$\pm$56& 205$\pm$151& 159$\pm$51\\
   solution 2 & 137$\pm$137 & 173$\pm$47 & 138$\pm$54& 187$\pm$150& 125$\pm$50\\
   \hline
 \end{tabular}
\caption{Stau masses preliminary determined from fitting of $\tau \tau$-edges. 
For each set two stau mass solutions are given.}\label{tab:massfitsu3}
\end{center}
\end{table}

It is seen from Table (\ref{tab:massfitsu3}) that uncertainties of stau mass
reconstruction are very large and additional steps in stau mass reconstruction
are required.

Eq.(\ref{endtau}) gives two solutions for stau mass. Because we use these masses
as a rough estimate for more precise procedures described below in the following
an average of these two solutions is used as an estimate for the stau mass.
Final results are not sensitive to this approximation. In particular, a good 
estimate for the stau mass at this stage would be just an average of two
lightest neutralino masses.

\section*{IV. Background suppression}

As a second step of stau mass reconstruction an event filter is used 
to suppress  background before the final fit.
At the event filter stage we assume that three light SUSY particle masses 
($\tilde \chi_{2}^{0}$, 
$\tilde \tau_{1}$, $\tilde \chi_{1}^{0}$)
are fixed. The mass of $\tilde \tau_{1}$ is taken as the preliminary mass found by the endpoint
method in the previous chapter. Neutralino masses are taken from Table \ref{tab:reco1}.
Gluino and sbottom masses are assumed
to be distributed uniformly in the range 
$m_{\tilde g} \pm 2\sigma$, $m_{\tilde b} \pm 2\sigma$ 
where masses and errors are given in Table \ref{tab:reco1}.

The event filter procedure is based on minimization for each event of  
the function 

\begin{equation}
\chi^2(m_{\tilde g}, m_{\tilde b}) =  \sum_{i=1}^4 \frac{(p_i^{event}-p_i^{meas})^2}{\sigma_{i}^2} + \lambda f(\vec m, \vec p)~
\label{chi1}
\end{equation}

where index i 
runs over 
two b-quarks and two
taus and labels the measured absolute momenta $p_{i}^{meas}$,  the uncertainties $\sigma_i$ 
in their measurement,
and the  event true absolute momenta $p_{i}^{event}$.
The approximate function (\ref{chi1}) takes into account only uncertainties in
tau and b-jet energy measurements.  
Note that positions of each of two
b-jets and of each of two taus in the decay chain (\ref{chain}) are unknown. 
It is quite simple to resolve the b-jets assignment because usually 
(about $96\%$ of the time) the b-jet 
with higher $p_T$ originates from the $\tilde b$-quark decay.
Therefore, for each event we assume that the b-jet 
with higher $p_T$ originates from the $\tilde b$-quark decay.
Taus with higher $p_T$ (after cuts) are produced in both vertices with comparable 
probabilities. In this work  it is assumed 
that taus with higher $p_T$ originate from $\tilde \chi_2$  decay.
We use the following parametrization for 
$\sigma_i$ in equation (\ref{chi1}): 
for b-jets and tau-jets ~ $\sigma/E = 0.5/\sqrt {E(GeV)} ~\oplus~ 0.03$. 
The Lagrange multiplier $\lambda$ in Eq.(\ref{chi1}) takes into account 
the mass relation constraint $f(\vec m, \vec p)$  \cite{tovey} - \cite{dk} 
which relates masses and momenta of particles in the chain (\ref{chain}).

For  signal  events, the event likelihood distribution  
has a maximum in the region of the ($\tilde g$, $\tilde b$) mass plane correlated with
the true masses of $\tilde g$ and $\tilde b$. 
Thus signal events should give a peak in the region of true masses. 
For background events there is no strong
correlation of maximum likelihood distribution with true ($\tilde g$, $\tilde b$) masses. 
Therefore if we chose arbitrary a point in the ($\tilde g$, $\tilde b$) mass plane
in the range $m_{\tilde g} \pm 2\sigma$, $m_{\tilde b} \pm 2\sigma$ $\chi^2$ reaches 
its minimum in this region with much higher probability for a signal event than
for a background one. 
For each event, $10^5$ points are generated randomly in the mass plane range  
$m_{\tilde g} \pm 2\sigma$, $m_{\tilde b} \pm 2\sigma$ and the $\chi^2$
is calculated. If $\chi^2 < 10$ in at least 1000 points this event is
considered as a signal candidate and it is retained for the
subsequent analysis.

After the application of the event filter the ratio of background
events to signal events is reduced approximately by a factor 1.5 as can be
seen from Table \ref{tab:efsu3}. The contribution of the combinatorial
background is given  approximately by ((Signal+Background)/Signal)$^5$,
corresponding to five event combinations required at the last step of
mass reconstruction,
and is therefore significantly suppressed. The suppression factor 
varies from 5 to 10. Note that any five event combination including
at least one background event is considered as a background.

\begin{table} [htb!]
\begin{center}
 \begin{tabular}{|c|c|c|c|c|c|}
   \hline
   Set &Total & Signal & SUSY &Ratio & Suppression\\
    &Number & Events & Backg. & & Factor\\
   \hline\hline
   1  & 199/98& 47/33 & 152/65 & 4.2/3.0 & 5\\
   \hline
   2  & 212/118& 56/49 & 156/69 & 3.8/2.4 & 10\\
   \hline
   3  & 216/116& 53/39 & 163/77 & 4.1/3.0 & 5\\
   \hline
   4  & 216/114& 48/37 & 168/77 & 4.5/3.1 & 6\\
   \hline
   5  & 229/124& 65/55 & 164/69 & 3.5/2.3 & 8\\
   \hline\hline
   1-5  & 1072/570& 269/213 & 803/357 & 4.0/2.7 & 7\\
   \hline
 \end{tabular}
\caption{The number of signal and background events before/after 
an application of event filter to 160k SUSY events. 
The last row shows the sum over  all five sets. 
Ratio = (Signal+Background)/Signal.
The last column
gives approximately a suppression factor (Ratio$^5$ before to 
Ratio$^5$ after) for five event background combinations.}
\label{tab:efsu3}
\end{center}
\end{table}

\section*{V.  Stau mass reconstruction}

As a third step a combinatorial procedure \cite{dk} is used for the final stau mass reconstruction.
It is applied only to the events that pass the event filter.
At the final stage of mass reconstruction when the physical
background has already been reduced, we will consider all possible
five event combinations from the event sample.

SUSY particle masses are reconstructed by a search  
for a maximum of a combined likelihood function constructed for each possible
combination of five events in the data sample. 

The $\chi^2$ function for an event is defined by

\begin{equation}
\chi^2_{event} = \sum_{i=1}^4 \frac{(p_i^{event}-p_i^{meas})^2}{\sigma_{i}^2} 
+ \sum_{n=1}^5 \frac{(m_n^{event}-m_n)^2}{\sigma_{n}^2}  + \lambda_1 f + \lambda_2 f^{ll} ~.
\label{chi2}
\end{equation}
where the first term takes into account deviations of measured
momenta of b-jets and tau-jets from the true ones. The second term
takes into account that the masses of sparticles vary from event to event 
as approximated by a Gaussian of width $\sigma_n$
instead a Breit-Wigner distribution.
In Eq.(\ref{chi2}) the mass relation and $\bf {\tau \tau}$ edge constraints are taking into account
by Lagrange multipliers $\lambda_{1}, \lambda_{2}$.
Standard deviations corresponding to the mass widths are taking to be 15 GeV for the gluino,
5 GeV for bottom squark and 1 GeV for light masses. 
The first two numbers are comparable with
theoretical widths for heavy SUSY particles. The last number takes into account
the fact that light SUSY particles are quite narrow or stable. We note 
that the results of the mass reconstruction are not strongly sensitive to the actual values   
of sparticle widths.

At this step it is assumed that the starting distributions of the 
heavy masses are uniform in the range: 
mean value $\pm ~2\sigma$,
where the mean values and standard deviations 
are given in Table \ref{tab:reco1}. For the three light masses a Gaussian distribution 
is assumed with mean values and standard deviations 
found by the endpoint technique as given in Table \ref{tab:massfitsu3}. 
The MINUIT code is used 
to search for the minimum of  the $\chi^2$ function for five event set with five mass parameters.

\begin{figure}[htb!]
  \centering
  \includegraphics[width=0.6\textwidth]{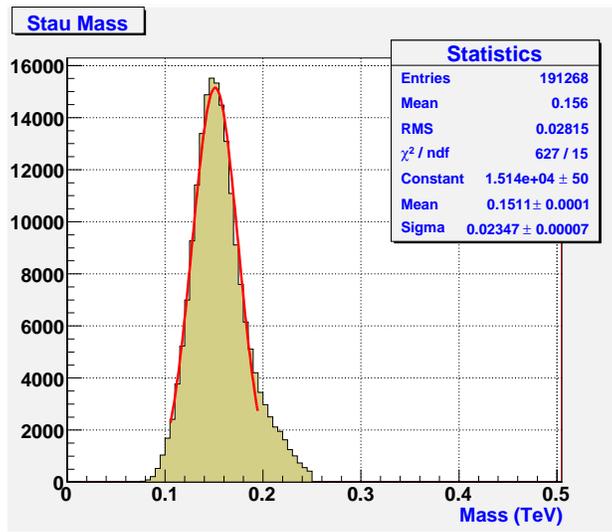} 
  \caption[Short caption.]{A reconstructed stau mass distribution
  for Set 1, including SUSY background with integrated luminosity 8 $\rm fb^{-1}$.
  The line is a result of Gaussian fit.   
   }
\label{fig_mass1}
\end{figure}

The reconstructed SUSY particle mass distribution and a result of fitting this distribution by a Gaussian are
presented in Fig.\ref{fig_mass1} for Set 1.  
As can be seen in this figure, the reconstructed mass distribution is well described by a Gaussian
except for small tails.
The mean values of the Gaussian fit are the reconstructed sparticle masses.

Final results  of this mass reconstruction approach 
are presented 
for five data sample sets of 160k events each 
in Table \ref{tab:reco}.

\begin{table} [htb!]
\begin{center}
 \begin{tabular}{|c|c|c|c|c|c|}
   \hline
   Particle & Set 1 & Set 2 & Set 3 & Set 4 & Set 5  \\
   \hline\hline
    $\tilde{\tau}_{1}$ & 154$\pm$24 & 192$\pm$23 & 168$\pm$25 & 202$\pm$29 & 150$\pm$22\\
   \hline
 \end{tabular}
\caption{Reconstructed stau masses and 
         reconstruction errors for five data sample sets of 160k events each.}
\label{tab:reco}
\end{center}
\end{table} 

It follows from Table \ref{tab:reco} that reconstruction errors after the final step are
significantly reduced in comparison with preliminary estimates \ref{tab:reco1} obtained by end-point
method.

\begin{figure}[htb!]
  \centering
  \includegraphics[width=0.6\textwidth]{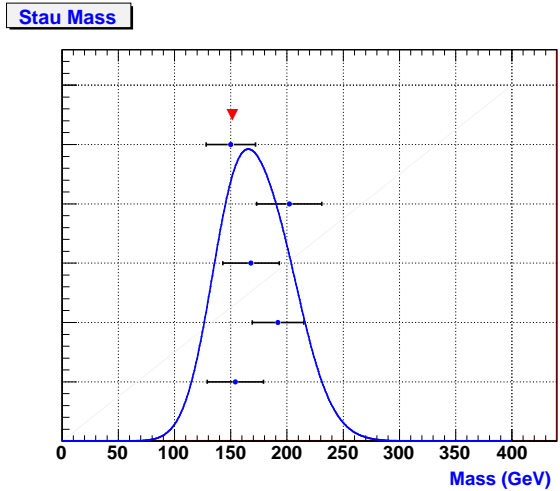} 
  \caption[Short caption.]{An ideogram of reconstructed stau masses, 
including SUSY background for five data sample sets with integrated luminosity 8 $\rm fb^{-1}$. 
The triangle marker gives the position of the SU3 theoretical mass.
Points with error bars correspond  to five data sample sets.}
\label{fig_ideo1}
\end{figure}

In order to illustrate a spread in reconstructed masses
the results of Table \ref{tab:reco} are also shown in a form of ideogram \cite{graph}
in Fig.\ref{fig_ideo1}
for the five data sample sets. 
Each reconstructed mass in an  ideogram is represented by a Gaussian with a central value $m_i$, error $\sigma_i$ and
area proportional to 1/$\sigma_i$. The solid curve is a sum of these Gaussians.

The Gaussian form of the ideogram and relatively small shift of peak positions
with respect to theoretical masses demonstrate the self-consistency of
stau mass reconstruction approach.

\section*{IX. Conclusion}

We applied the cascade mass reconstruction approach developed in \cite{dk} 
to reconstruction of $\tilde{\tau}_{1}$ mass at the LHC 
with a low integrated luminosity of about $8 ~\rm fb^{-1}$ at 14 TeV. 
This luminosity can be reached in the early stage of LHC operation.
At low integrated luminosity stau mass reconstruction is complicated 
because escaping neutrinos modify the edge of the $\tau \tau$ invariant mass distribution,
reconstructed stau mass is very
sensitive to uncertainties in input parameters (neutralino masses) and the 
level of SUSY background is high.

Our approach to the stau mass reconstruction was based on a 
consecutive use of the endpoint method, 
an event filter and a combinatorial mass reconstruction method. 

The endpoint method allows  
preliminary estimate of $\tilde \tau_{1}$ mass by 
assuming that sparticle masses required as input 
parameters for this procedure are known from the same experiment
with a precision of about $10\%$ for heavy sparticles (such
as gluino and bottom squark)
and  about $20\%$  for the two lightest neutralinos.
In this work these input masses were obtained by the cascade mass reconstruction
approach \cite{dk}. At high integrated luminosity of $100-300 ~\rm fb^{-1}$
they could be found by using one of techniques
proposed in \cite{baer} - \cite{nojiri}.

The stau mass was reconstructed with a precision of about $20\%$

\section*{Acknowledgments}  
The authors thank K.Cranmer, M.Ibe, A.Mincer and P.Nemethy for interesting discussions and useful suggestions.
This work has been supported by the National Science Foundation under grants PHY 0428662, PHY 0514425
and PHY 0629419.

\end{document}